# Unusual Oscillation in Tunneling Magnetoresistance near a Quantum Critical Point in $Sr_3Ru_2O_7$

Joe Hooper and Zhiqiang Mao[1]*

Robin Perry and Yoshiteru Maeno[2]

[1]*Physics Department, Tulane University, New Orleans, Louisiana 70118, USA.*

[2]*Department of Physics, Kyoto University, Kyoto 606-8502, Japan*

PACS numbers: 75.30.Kz, 71.27.+a, 73.40.Gk

We performed single electron tunneling measurements on bilayer ruthenate $Sr_3Ru_2O_7$ under magnetic fields at low temperature. We observe an unusual oscillation in tunneling magnetoresistance near the metamagnetic quantum phase transition at temperatures below 7 K. The characteristic features of this oscillation suggest that it is unrelated to traditional quantum oscillations caused by orbit quantization. In addition, tunneling spectra are found to change sharply in the low bias voltage range of $V < 2mV$ near the transition field. These observations reveal that the Fermi surface of $Sr_3Ru_2O_7$ changes in a surprising way as the system undergoes strong critical fluctuations.

Quantum criticality is among the most actively researched topics in contemporary condensed matter physics. An understanding of quantum criticality and non-Fermi liquid behaviour near quantum critical points (QCPs) is fundamental to the physics of strongly correlated electron systems [1-3]. In general, QCPs are reached by tuning the characteristic temperature of a second-order phase transition to absolute zero using a control parameter such as chemical composition or pressure [1]. Recently, a new class of field-tuned QCP, a metamagnetic critical-end point, was observed in bilayer ruthenate $Sr_3Ru_2O_7$ [4]. The quantum phase transition (QPT) associated with this QCP is a first-order transition in which no symmetry breaking is involved, fundamentally different from QPTs produced by depressing a second-order phase transition temperature to absolute zero [4-6].

The strontium ruthenates of the Ruddlesden-Popper series $Sr_{n+1}Ru_nO_{3n+1}$ display a remarkable range of unique superconducting and magnetic properties [7,8]. $Sr_3Ru_2O_7$, the $n=2$ member of the series, has an intermediate dimensionality between the compounds with $n=1$ ($Sr_2RuO_4$, a spin triplet superconductor [7]) and $n=\infty$ ($SrRuO_3$, an itinerant ferromagnet [9]). At zero field the ground state of $Sr_3Ru_2O_7$ is a paramagnetic Fermi liquid very close to a ferromagnetic instability [10]. The metamagnetic QPT occurs at moderate applied fields, 5-6 T for $H//ab$ and ~7.7 T for $H//c$ [6]. Neutron scattering studies on this material reveal no long range ordering but do indicate weak antiferromagnetic (AFM) correlations at low temperatures [11].

In general, an itinerant metamagnetic transition is ascribed to a field-induced non-linear exchange spin splitting of the Fermi surface (FS) [12,13], resulting in a change of FS topography across the transition. De Haas van Alphen (dHvA) and Shubnikov-de Haas (SdH) measurements have provided unambiguous evidence for this picture in heavy fermion compounds such as $UPt_3$ [14] and $CeRu_2Si_2$ [15,16]. The metamagnetism in $Sr_3Ru_2O_7$ is expected to have a similar mechanism [5,6]. However, the FS of $Sr_3Ru_2O_7$ above and below the transition field has not yet been fully mapped out by either dHvA or SdH measurements.

Here, we present experimental data obtained from single-electron tunneling measurements on $Sr_3Ru_2O_7$, which reveal that the FS changes in a surprising way as this material undergoes the metamagnetic transition. We found that the tunneling magnetoresistance exhibits unusual oscillations with the change of FS.

Tunneling is a well known quantum mechanical process whereby electrons can penetrate through a barrier between two materials. Tunneling can occur through a thin insulating layer between a normal metal and a superconductor (N-I-S), between two superconductors (S-I-S), or between two normal materials (N-I-N). While tunneling experiments on N-I-S and S-I-S junctions have been extensively used to investigate gap structures of superconductors, tunneling in N-I-N junctions is useful for studying the properties of the FS in non-superconducting materials. The general expression for tunneling current can be written as

$$I = (2\pi eA/\hbar) | H_T |^2 \int N_1(E-eV) N_2(E) [f(E-eV)-f(E)] \, dE, \qquad (1),$$

where $A$ is the area of the junction, $H_T$ is the tunneling matrix element (assumed to independent of the energy near $E = 0$), $V$ is the bias voltage, $N_1(E-eV)$ and $N_2(E)$ are the DOS for the two materials separated by the insulating barrier, and $f(E-eV)$ and $f(E)$ are Fermi distribution functions [17]. For N-I-N type junctions, Eq. (1) can be approximated as

$$I \approx c \, N_1(\varepsilon_F) N_2(\varepsilon_F) \, e \, V, \qquad (2),$$

where $c$ is a constant, and $\varepsilon_F$ is the Fermi energy. From this formula, it can be seen that tunneling conductance can provide information on the DOS at the FS of one material if the other side has a known FS. Our tunneling measurements were performed on N-I-N type junctions prepared on high-quality single crystals of $Sr_3Ru_2O_7$.

Crystals for these experiments were grown crucible-free in an image furnace at Kyoto University. X-ray diffraction analysis on the crystals used for this study did not reveal any impurity phase. Resistivity measurements showed residual resistivities of ~1-2 $\mu\Omega$ cm, comparable to previously reported high quality samples [6]. Tunneling junctions of $Sr_3Ru_2O_7$ - I - Au were prepared on finely polished (100) faces (see the inset of Fig. 1). Tunneling measurements were carried out in a $^3$He cryostat (equipped with a 9 T magnet) using a standard four-probe technique.

Figure 1 shows the resistance of a typical $Sr_3Ru_2O_7$ - I - Au junction as a function of magnetic field $R_J(H)$ ($H//c$), measured at $T = 0.3$ K. First, we focus on the data above 2 T. The most striking feature above this field is a remarkable oscillation of junction resistance which develops from ~3 T and grows considerably in amplitude for fields above 5 T. The maximum amplitude observed below 9 T corresponds to a 2.5% change in $R_J(H)$. The background of junction resistance (ignoring the oscillations) decreases with increasing applied field. This downward trend is also greatly enhanced above 5 T, where the oscillation amplitude begins to increase sharply. At fields below 2 T, another prominent feature was observed. Small amplitude oscillations first appear around 1 T, followed by an extremely sharp jump at 1.56T. All these phenomena were consistently reproduced over a range of samples.

For comparison, similar tunneling measurements were performed on pure $Sr_2RuO_4$ crystals (shown in Fig.1); no oscillations were observed and junction resistance shows a slightly positive magnetoresistance, in sharp contrast to $Sr_3Ru_2O_7$ junctions. Therefore, the oscillation phenomenon shown in Fig.1 should be an intrinsic property of $Sr_3Ru_2O_7$. Since the DOS at the FS in Au is approximately a constant, from Eq. 2 it appears that oscillations in junction resistance reflect an oscillation in the DOS at the FS of $Sr_3Ru_2O_7$. However, we note that no corresponding oscillation feature was observed in bulk resistivity or susceptibility measurements of $Sr_3Ru_2O_7$ [6]. Hence it may be premature to simply attribute the oscillation of tunneling magnetoresistance observed in our experiment to the oscillation of DOS at the FS (more discussions are given below). Additionally, we would like to point out that although the oscillation does not show a discontinuous change at the metamagnetic transition field, the tunneling spectra do change significantly around this point (see below), consistent with the observation of metamagnetism in bulk measurements [4,6].

The temperature dependence of $R_J(H)$, displayed in Fig. 2, reveals a systematic reduction in the amplitude of oscillations with increasing temperature. By 7 K, the oscillations become almost unobservable and $R_J(H)$ shows an approximately linear negative magnetoresistance. The positions of oscillation maxima and minima are found to be temperature dependent. The discontinuity at 1.56T, as well as the small oscillations at low fields, also show a systematic reduction. In addition, we observed an interesting hysteresis behavior in $R_J(H)$ between upwards (dotted curves) and downwards field sweeps (solid curves). This hysteresis is significantly enhanced above 5 T; it is also temperature dependent, disappearing above 3 K.

The dHvA / SdH effects are well established means of determining the Fermi surface via characteristic oscillations occurring when Landau levels cross the Fermi level. Our data, while also showing oscillations under applied fields, appears to be unrelated to dHvA/SdH-type phenomenon for the following reasons. First, we did not observe any oscillation in bulk resistivity measurements on a crystal with a residual resistivity approximately the same as the crystals used for preparing junctions. Second, SdH oscillations recently observed in ultra-pure crystals of $Sr_3Ru_2O_7$ with a residual resistivity of 0.3 $\mu\Omega$.cm [18] show distinctly different features from our observations. Fourier analysis of these SdH oscillations in the field range 4-7 T yields two frequencies, 140 T and 1800 T; in contrast, frequencies derived from our oscillations are strongly field dependent. Fourier analysis of the 5-9 T range yields a single frequency of 0.7 T, while the 3-5 T range yields a frequency of 50 T. These frequencies did not change between junctions made along different in-plane directions; only changes in relative amplitude were observed.

Although these features strongly suggest that the oscillation in tunneling magnetoresistance is not a SdH-type phenomena, it is worthwhile to consider what conclusions might be drawn if we assume that these *are* related to standard quantum oscillations. We have estimated the wave vector $k_F = (2\pi eF/\hbar\pi)^{1/2}$ (*F*: oscillation frequency) using the frequencies derived above. The frequencies of 50T and 0.7T correspond to a $k_F$ of 0.039 Å$^{-1}$ and 0.00461 Å$^{-1}$, respectively. The Fermi velocities estimated from these wave vectors are $4.5 \times 10^4$ m/s and $5.0 \times 10^3$ m/s. The former appears to be comparable to the average Fermi velocity ($1.3 \times 10^5$ m/s) of the lens-shaped band predicted in band structure calculations [19]. However, the latter one seems too small to fit any predicted bands. Thus, our data would imply that only a very tiny portion of the Fermi surface participates in the tunneling process. This seems less likely considering the orthorhombic structure of $Sr_3Ru_2O_7$ [19,20] and the fact that we did not observe any oscillation frequency change between junctions along different in-plane directions. This further supports our above argument that that oscillation reported here is a unique property of $Sr_3Ru_2O_7$, unrelated to traditional dHvA/SdH oscillations.

In addition, we note that the background trend of junction resistance decreases with increasing field (see Fig.1), in contrast to bulk magnetoresistance which increases with field [6]. This divergent behavior stems from the fact that these two different techniques probe different properties. Junction resistance is primarily determined by the DOS near the FS (as shown in Eq. 2), while bulk resistivity is dominated by the scattering mechanism. The decreasing trend in junction magnetoresistance indicates that the DOS increases with increasing field, while the positive magnetoresistance seen in bulk measurements is likely related to AFM correlations [11].

To further investigate how the oscillation in $R_J(H)$ is related to critical fluctuations as the system approaches the quantum metamagnetic transition, we measured the tunneling spectra of $Sr_3Ru_2O_7$ - I - Au junctions at different fields and a fixed temperature of 0.3 K. Figure 3 shows the tunneling spectra of $dI/dV$ (denoted by $G$ below) taken at 0, 5.98, 6.48, 8.15, and 9 T. At zero field the tunneling spectrum shows a sharp cusp around zero bias, indicating a suppression of DOS at the FS of $Sr_3Ru_2O_7$. This is likely associated with AFM correlations caused by FS nesting [11,19]. With increasing applied field, the sharp cusp becomes suppressed. When field is raised to about 8 T (close to the QCP of $Sr_3Ru_2O_7$), the spectrum is flattened considerably around zero bias. However, when field is increased past the critical point, the sharp cusp redevelops. Additionally, we note that the significant changes in the tunneling spectra under field occur only within the low voltage range $V < 2$ mV. This is consistent with the low characteristic energy of AFM spin fluctuations ($2.3 \pm 0.3$ meV) determined by neutron diffraction [11].

To better delineate these spectrum changes under magnetic field, we attempted to fit these data to a power-law bias dependence, $G = G_0 + AV^n$. We found that this fit is good within a bias range of $0.5 < V < 2.0$ mV as shown in the inset of Fig. 3, but has a larger error in the lower bias-range of $V < 0.5$ mV. Figure 4b shows the field dependence of the exponent $n$ obtained from fitting the spectra at different fields. $n$ increases almost linearly below 5 T, more rapidly above 5 T, then sharply near the metamagnetic transition; it reaches a maximum at about 8 T and drops dramatically when the field is past the critical field. $n$ can be viewed as a characteristic parameter describing the FS since it is closely associated with the energy dependence of the DOS. Consequently, the sharp change of $n$ around 8 T can be considered strong evidence for a sharp change of FS near the metamagnetic transition.

Since tunneling resistance strongly depends on the properties of the FS as shown in Eq. 1, the oscillation in tunneling magnetoresistance observed in our experiment, which may not directly reflect an oscillation of DOS, does suggest that the field-induced change of FS is surprising. The fact that both the oscillation amplitude and the exponent *n* increase prominently above the same field, 5 T (see Fig. 4), further supports the notion that they are closely related. In terms of the existing theory of metamagnetism in $Sr_3Ru_2O_7$, the change of FS near the metamagnetic transition originates from a non-linear exchange-enhanced spin splitting of the FS [5]. The remarkable oscillation we observed is unexpected within the current picture. According to that theory, $Sr_3Ru_2O_7$ appears to be an ideal material for testing the original model of quantum critical phenomenon [12,21], and thus our experiment raises new interesting questions about the underlying physics.

Transport and thermodynamic measurements have revealed strong quantum critical fluctuations near the QCP in $Sr_3Ru_2O_7$ [6], as mentioned above. These critical fluctuations can be pictured as FS fluctuations [5,6]. The non-Fermi liquid behavior at the critical point can be attributed to the strong fluctuation of the position of the spin-up and spin-down Fermi surfaces [5]. The oscillation in tunneling magnetoresistance discussed above appears to be closely related to these strong critical fluctuations, since the oscillation amplitude and hysteresis are enhanced significantly as the system is tuned close to the critical point. Due to the presence of these strong critical fluctuations, the function *f(E)* or *f(E-eV)* in the integrals of Eq. 1 is likely to substantially deviate from the Fermi distribution function as the system approaches the critical field. Eq. 2, therefore, may not be valid near the transition field. Incorporating strong quantum critical fluctuations into tunneling processes seems necessary to develop a theory for this new oscillation phenomenon.

In summary, near the metamagnetic transition in $Sr_3Ru_2O_7$ we have observed an unusual oscillation in tunneling magnetoresistance and a sharp change of the FS in tunneling spectra. The amplitude of this oscillation increases significantly near the transition field, thus suggesting that this oscillation is linked with critical fluctuations of the FS. This phenomenon is unexpected within the current theoretical model of metamagnetism for $Sr_3Ru_2O_7$, and therefore raises new interesting questions about the underlying physics. Further studies to determine if this phenomenon is universal to other itinerant metamagnetic materials are certainly of great interest. Finally, we would like to point out that single-electron tunneling might prove to be a very effective means of studying quantum critical phenomena.

The authors would like to thank A. P. Mackenzie, Y. Liu, J. Zhu, and Q. Si for valuable discussions. This work was supported by the Louisiana Board of Regents support fund through Grant No. LEQSF(2003-6)-RD-A-26, Tulane's Committee on Research Summer Fellowship, and by the NSF's East Asia Summer Institutes for Graduate Students.

*Correspondence should be addressed to zmao@tulane.edu.

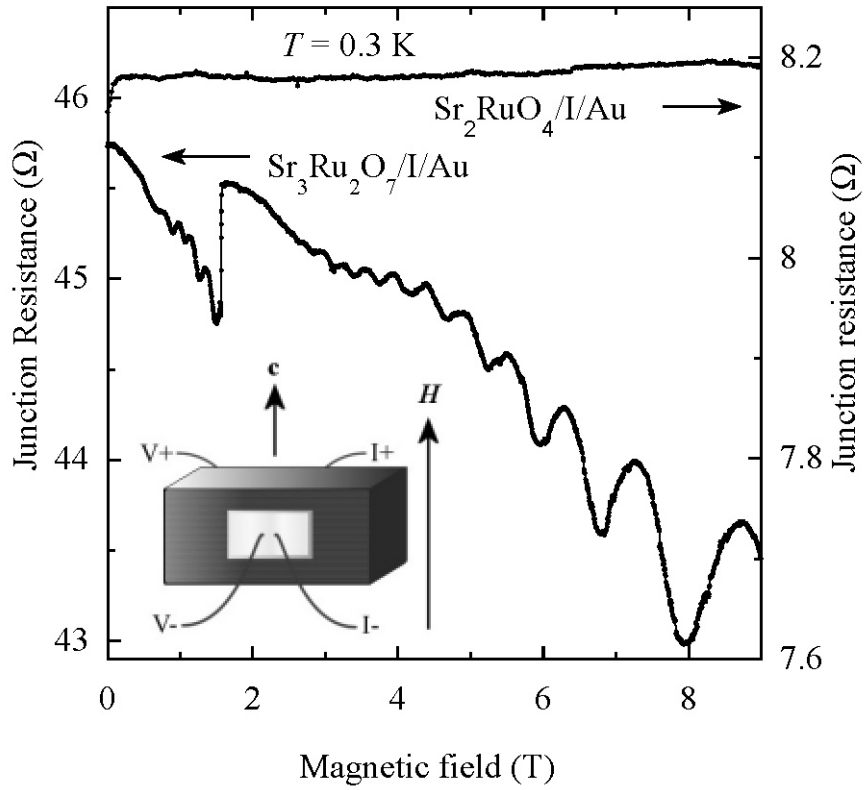

FIG 1: Resistance of a typical $Sr_3Ru_2O_7$ - I - Au junction versus applied field $R_J(H)$ ($H//c$). The inset shows the experimental junction setup, with a 60Å insulating barrier of $Al_2O_3$ and a 500 Å Au counterelectrode. The $Al_2O_3$ layer was prepared by three sequential evaporations of 20 Å of Al in high vacuum, each followed by oxidation in air. The gold layer was then evaporated on top of the $Al_2O_3$. $R_J(H)$ was measured with a current of 4.8 μA, corresponding to a bias of 0.22 mV. For comparison, a similar measurement on $Sr_2RuO_4$ is shown; the slight drop at ~0.065T is due to the superconducting transition.

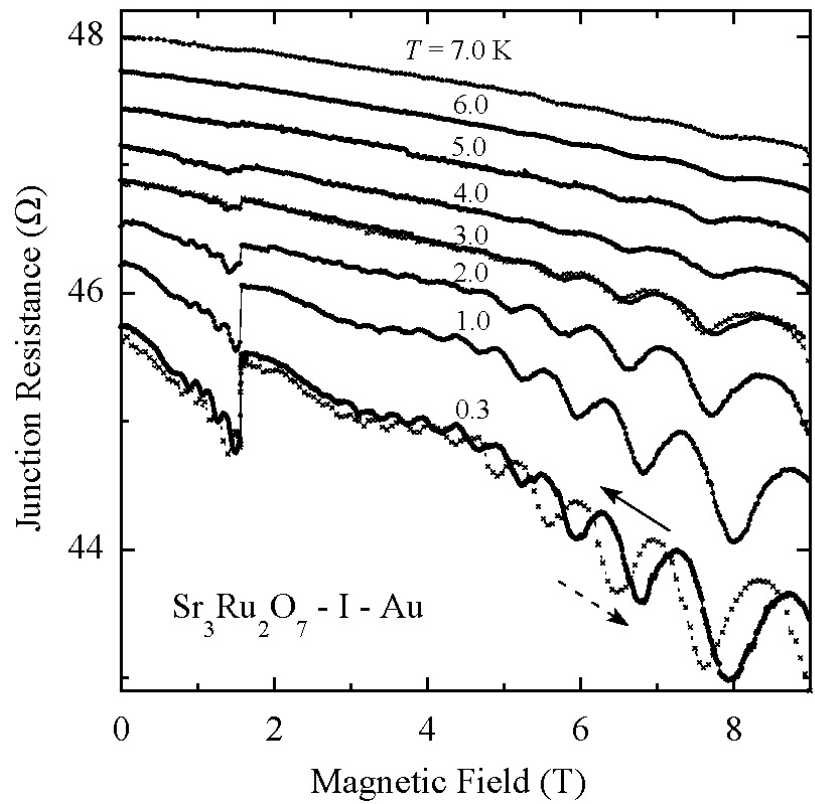

FIG 2: Temperature dependence of $R_J(H)$ of a $Sr_3Ru_2O_7$ - I - Au junction. All plots except $T = 0.3$ K are shifted for easier comparison. Solid black curves are taken with field sweeping down; the dotted curves ($T = 0.3$ K and 3 K) represent field sweeping up.

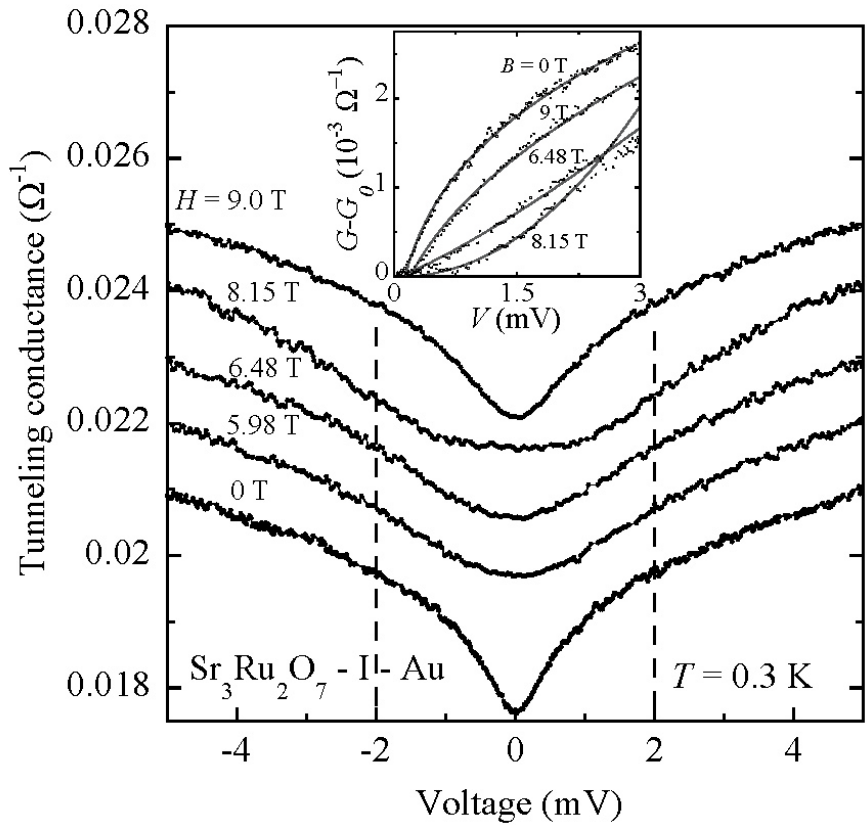

FIG 3: Tunneling conductance d$I$/d$V$ as a function of bias voltage at constant applied fields and temperature. The tunneling spectra only change significantly with applied field within a low bias range (< 2 mV, denoted by the dotted lines). All curves except $H$ = 9 T are shifted for easier comparison. The inset shows tunneling conductance normalized to zero; solid lines show a power-law fit ($G - G_0 \propto V^n$). $n$ is the exponent obtained from this fitting; the values of $n$ are 0.63, 1.18, 1.21, 2.12, and 0.52 for fields of 0 T, 5.98 T, 6.48 T, 8.15 T, and 9 T respectively.

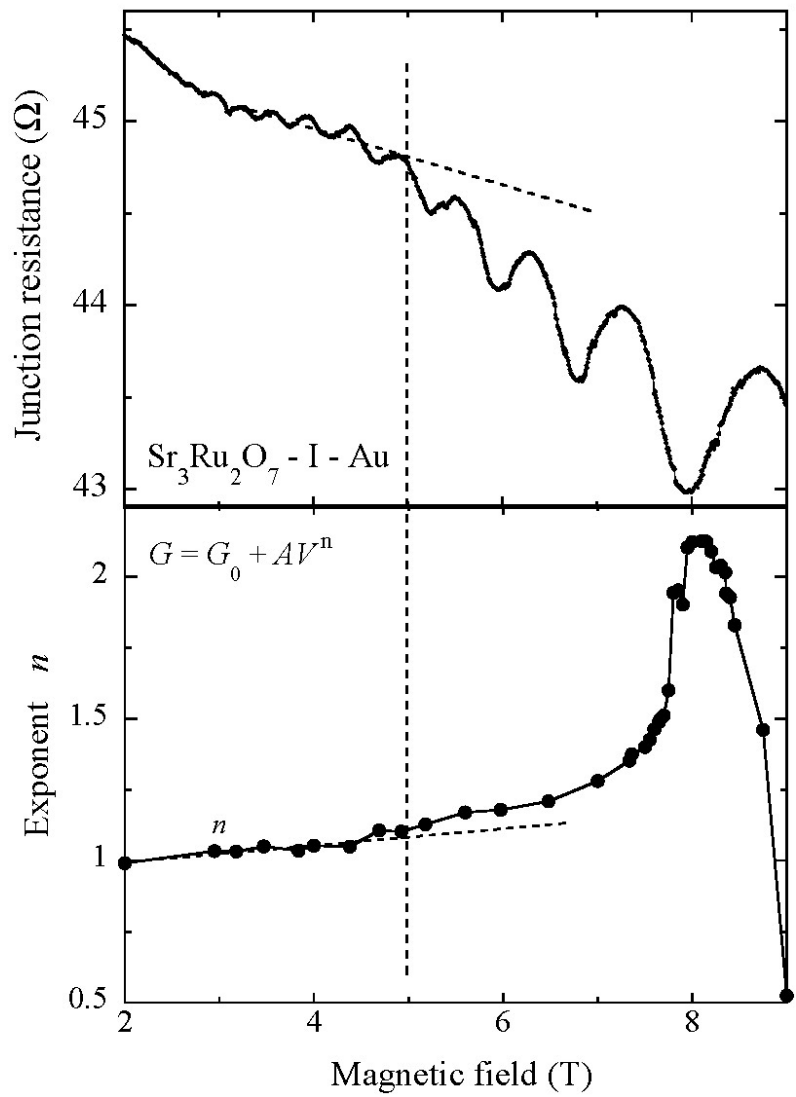

FIG 4: (a) Field dependence of junction resistance (taken with field sweeping down); (b) Field dependence of $n$, the exponent of the power-law fit of tunneling conductance ($G = G_0 + AV^n$). The dashed lines are a guide to the eyes to show that both the oscillation amplitude and the exponent $n$ increase prominently above 5 T.